# Quantization of Visible Light by a $Ni_2$ Molecular Optical Resonator

Miao Meng, Ying Ning Tan, Yu Li Zhou, Zi Cong He, Zi Hao Zhong, Jia Zhou, Guang Yuan Zhu, Chun Y. Liu*

Department of Chemistry, College of Chemistry and Materials Science, Jinan University, 601 Huang–Pu Avenue West, Guangzhou 510632, China

* Correspondence: tcyliu@jnu.edu.cn

**Abstract**

The quantization of an optical field is a frontier in quantum[1,2], and polaritonic chemistry,[3] with implications for both fundamental science and technological applications. Here, we show that a dinickel complex ($Ni_2$) traps and quantizes classical visible light, behaving as an individual quantum system or the Jaynes-Cummings (JC) molecule.[4,5] The composite system forms through coherently coupling the two-level Ni↔Ni charge transfer transition ($\omega_0$) with the local scattering field ($\omega_L$) at the single molecule level, which produces nonclassical light featuring photon anti-bunching[6] and squeezed states[7], as verified by a sequence of discrete photonic modes in the resonance fluorescence with weak incoherent excitation. in particular, in this $Ni_2$ system, the collective coupling of N-molecule ensembles, acting as an effective JC molecular system, scales as $N\sqrt{N}\Omega$, distinct from the Tavis-Cummings (TC) model[8], which allows easy achievement of ultrastrong coupling[9]. This is exemplified by a vacuum Rabi splitting of 1.2 eV of the resonance ($\omega_0$ = 3.25 eV or 382 nm) and a normalized coupling rate of 0.18 for the N = 4 ensemble. The resulting quantum light of single photonic modes enables driving the interaction of the molecules with the vacuum field in cavity-free solution, which profoundly modifies the electronic states. Our results demonstrate $Ni_2$ as a robust platform for quantum optical phenomena under ambient conditions, offering emergent pathways for molecular physics, polaritonic chemistry and quantum information processing.

## Introduction

The coherent coupling of a two-level excitation of an emitter (atom or molecule) with transition energy $\omega_0$, where the ground state is denoted as $|g\rangle$ and the excited state as $|e\rangle$, to a photonic state $|n\rangle$ of a single mode of light with frequency $\omega_L$, creates a hybrid system with entangled quantum states $|n, \pm\rangle$ (Figure 1A, left). This coupling profoundly modifies the electronic structure of the emitter. According to the Jaynes-Cummings (JC) model (Eq. 1 when $\hbar = 1$): [1,4,5,10,11]

$$H_{JC} = \omega_L a^+ a + \omega_0 \sigma_z + g(a\sigma_+ + a^+\sigma_-) \qquad (1)$$

where $a^+$ and $a$ are the photonic creation and annihilation operators, and $\sigma_+$ and $\sigma_-$ are the raising and lowering operators for the two-level system, and $\sigma_z$ represents the eigenstates of the excitation of the system, modeled as a spin-1/2 in a magnetic field. The coupling rate g parametrically quantifies the interaction strength and is proportional to the transition dipole moment ($\mu$) of the emitter and the electric field strength ($\boldsymbol{E}_0$), [1,4,5]

$$g = -\mu \boldsymbol{E}_0/\hbar \propto \mu/\sqrt{V} \qquad (2)$$

where $V$ is the effective mode volume. The eigenvalues of this coupled system form a ladder structure with periodic progressive eigenstates with increasing the level of the photon number state, known as the JC ladder (Figure 1A, middle),[11,12] which can be probed spectroscopically *via* features like the Mollow triplet[13,14] and Rabi doublet (Figure 1A, right).[15]

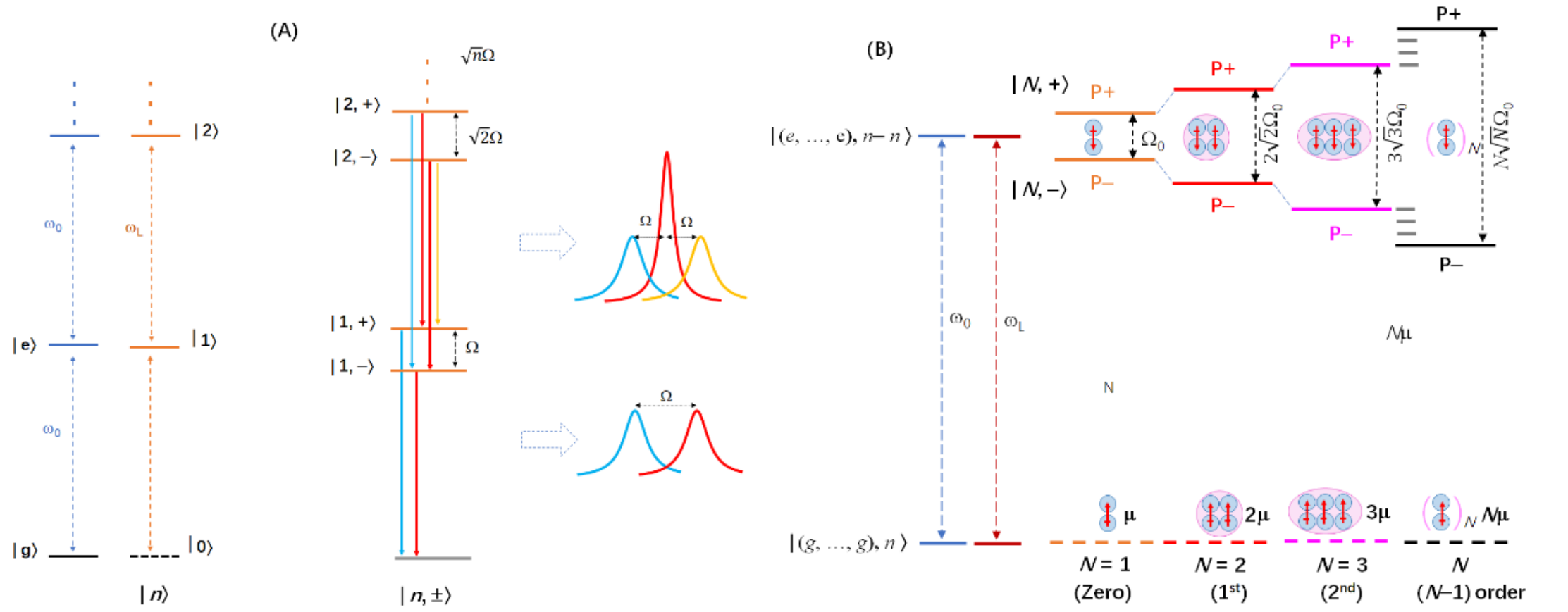

**Figure 1. Modifications of the eigenstates of a molecule by coupling the two-level excitation to an electromagnetic field.** (A) Resonant coupling between a two-level

excitation and photon number states (|$n$⟩) results in the formation of the Rabi doublet (lower) and the Mollow triplet (upper) in the resonance fluorescence spectra. The eigenstates are depicted as ladders, illustrating the progression with increasing the photon number states, a hallmark of strong light-matter interactions in the JC model. (B) Schematic illustration of the collective coupling effect in the $Ni_2$ molecular system, showing broadening of the vacuum Rabi splitting driven by simultaneous excitation of $N$ molecules in an ensemble with a total dipole moment $N\mu$ and by absorption of $N$ photons at resonance. The collective coupling strength scales as $N\sqrt{N}\Omega$, deviating from the conventional TC model, where the coupling scales as $N\Omega$.

Achieving strong or ultrastrong coupling requires either reducing $V$ to enhance $\boldsymbol{E}_0$ , or collectively coupling multiple emitters to a single mode of the electromagnetic field, [1,16] as described by the Tavis-Cummings (TC) model.[8] The collective coupling scales the interaction strength as $\sqrt{N}\,\Omega_0$, where $\Omega_0 = 2g$,[1,8,16] thus, effectively enhancing the coupling strength through the participation of $N$ emitters (see the Supplementary Material).

The JC model provides the theoretical framework for coupling a two-level matter system (e.g., atom, molecule, or quantum dot) to a single photonic mode.[1,4,10,11] When extended to an ensemble of $N$ identical JC molecules that are simultaneously but individually excited, the system becomes an effective "JC molecular system." For this ensemble, the Hamiltonian ($H_{JCN}$) can be expressed as:

$$H_{JCN} = N\omega_L a^+a + \omega_0 J_z + g(aJ_+ + a^+J_-) \tag{3}$$

where the collective operators $J_Z$ , $J_\pm$ are defined as the sums of the corresponding individual spin operators:[1,16,17] $J_z = \sum_{j=1}^{N} \sigma_{z,j}$, $J_\pm = \sum_{j=1}^{N} \sigma_{\pm,j}$, where $\sigma_{z,j}$ and $\sigma_{\pm,j}$ are the Pauli operators for the $j$-th molecule, acting on its individual Hilbert space. Note that Eq. 3 differs from the Tavis-Cummings model[1,8,16] with the $N$ factor in front of photon energy term because in this case, the $N$ JC molecules are simultaneously and individually coupled to $N$ photons. Although these individual operators are not numerically identical across molecules, they have the same form, reflecting the identical internal structure and uniform coupling of each molecule in the ensemble. Under such conditions, the ensemble behaves effectively as: $J_z = N\sigma_z$, $J_\pm = N\sigma_\pm$, where $\sigma_z$ and $\sigma_\pm$ denote the common operator forms of individual molecules due to the indistinguishability of the $N$ molecules. Rewriting Eq. 3 in terms of the spin operators for the single JC molecule gives:

$$H_{JCN} = N\omega_L a^+ a + N\omega_0 \sigma_z + Ng(a\sigma_+ + a^+\sigma_-) \quad (4)$$

Eq. 4 can be further decomposed as:

$$H_{JCN} = (N-1)\omega_L a^+ a + \{\omega_L a^+ a + N[\omega_0 \sigma_z + g(a\sigma_+ + a^+\sigma_-)]\} \quad (5)$$

For $N$=1, Eq. 5 reduces to the standard JC model (Eq. 1), yielding vacuum Rabi splitting $\Omega_0$.[11,16] For an ensemble with $N \geq 2$, the second term of Eq.5 describes the collective coupling of the $N$ individual emitters with a total dipole moment $N\mu$ (Figure 1B) to the single-mode (single photon) field, giving the coupling strength $\Omega = N\Omega_0$ according to Eq.2. Note that here the second term does not yield $\Omega = \sqrt{N}\Omega_0$, although it takes the form of the TC model (see the Supplementary Material), because it does not involve the Dicke states in the excitation of the $N$ emitters.[1,8,18] The first term accounts for addition of ($N$−1) photons to the molecular system so that overall to realize individual coupling of the $N$ emitters to the $N$ photons (Figure 1B). Integrating the first and the second terms in Eq. 5 introduces a factor $\sqrt{N}$ to the collective coupling strength $\Omega$, resulting in the “N-emitter and N-photon Rabi frequency”:

$$\Omega_N = \sqrt{(N-1)+1}\,\Omega = N\sqrt{N}\,\Omega_0 \quad (6)$$

To date, this $N\sqrt{N}\Omega$ collective coupling scaling for the JC multi-emitter system, which breaks down the dipole blockade, has not been experimentally demonstrated.

Optical cavities,[1,2,3,9,15] which enhance EM fields by confining photonic modes for strong light-matter interactions, have reached the sub-nanometer dimensions through construction of plasmonic cavities using spaced metal nanoparticles. By progressively shrinking the cavity size−for instance, using tip-to-tip plasmonic waveguides, surface plasmon field confinement approaches the quantum limit, approximately $10^{-8}\lambda^3$ (0.1 - 10 nm$^3$ for visible wavelength).[19] Recent studies show that when the junction gap decreases to $d$ < 5 nm, field-induced charge transfer (CT) occurs between the two metallic spheres, resulting in plasmonic charge transfer (PCT).[19, 20] As d further decreases to the single-digit angstrom range (< 1 nm),[21] the system enters the quantum tunneling regime and the CT transition undergoes a blue shift,[20] with the local EM field dramatically enhanced due to the extremely small mode volume.[21] Using the nanoparticle-on-mirror (NPoM) set-up, a space of 0.9 nm gap between the gold sphere and a golden mirror produces a PCT mode at 660 nm, which couples resonantly with a dye molecule's two-level transition at 655 nm.[22] However, such "naked" atom-scale cavities are unstable at room temperature due to their dynamic nature and potential laser-caused disassembly.[21] In the contact regime ($d \leq 0$) where quantum tunneling CT occurs, atomic wavefunction overlap results in covalent bonding between atoms,[20] creating a molecular system. This suggests that a well-designed molecular system can enable optical experiments at the quantum limit. In this context, studying the optical responses of complex molecules with a dimetal ($M_2$) unit acting as an atomistic resonator opens emergent avenues for polaritonic chemistry, molecular physics, and quantum optics.

Here, we demonstrate that a dinickel complex, $Ni_2(DAniF)_4$ (DAniF = di-(*p*-anisyl)formamidinate), acts as an atomistic resonator capable of quantizing visible light under ambient conditions, lowering the "cavity" dimensions to the atomic limit. Unlike metal nanoparticles, the $Ni_2$ molecular system exhibits distinct quantum optical responses due to the coherent coupling of the two-level Ni↔Ni CT transition and the local scattering field. Our results indicate that this hybrid system can trap classical light in the visible region and generate quantized states of light, characterized by a sequence of discrete photonic modes. The $Ni_2$ resonator operates without a traditional optical cavity, permitting exploration of strong, even ultrastrong light-matter interactions at ambient conditions using popular instrumentations. Furthermore, it allows for the investigation of the collective coupling of *N*-molecule ensembles, which significantly diverges from the TC model, showcasing collective coupling that scales as $N\sqrt{N}\Omega_0$. These properties establish the $Ni_2$ molecular system as a new platform for future research in polaritonic chemistry,[3,23] molecular qubits[2,24], and other applications[1,25], in quantum optics under ambient conditions.

## Results and Discussion

**Molecular Structure and the $Ni_2$ Geometry.** Following our primary study on the dimolybdenum complexes,[26] we now investigate the dinickel ($Ni_2$) complex $Ni_2(DAniF)_4$, coordinated by four DAniF ligands (Figure 2A). The molecular and electronic structures, as well as the absorption spectra of $Ni_2(DAniF)_4$ have been studied from a chemical perspective.[27,28] Formally, these dinickel formamidinates are considered to have an unbound $Ni_2$ unit with two $Ni^{2+}$ ions in a square planar coordination geometry (Figure 2A). The Ni···Ni nonbonding distance, as determined by X-ray diffraction, is 2.48(5) Å.[27] The $Ni_2$ unit can be described by a cylindrical model (Figure 2A, inset), with two Ni atoms separated by 0.248 nm ($D$) and an ionic ($Ni^{2+}$) radius of 0.063 nm ($R$),[29] giving an effective interfacial distance $d$ = 0.122 nm. The free space volume between the two Ni atoms is then determined to be ~ $10^{-3}$ $nm^3$, an order of magnitude smaller than the reported minimum volume of the plasmonic picocavity.[21] Therefore, the $Ni_2$ complex, acting as an atomistic optical resonator, enables extremely tight confinement of fields, facilitating the quantization of visible light under ambient conditions, even without the need for a traditional cavity setup.

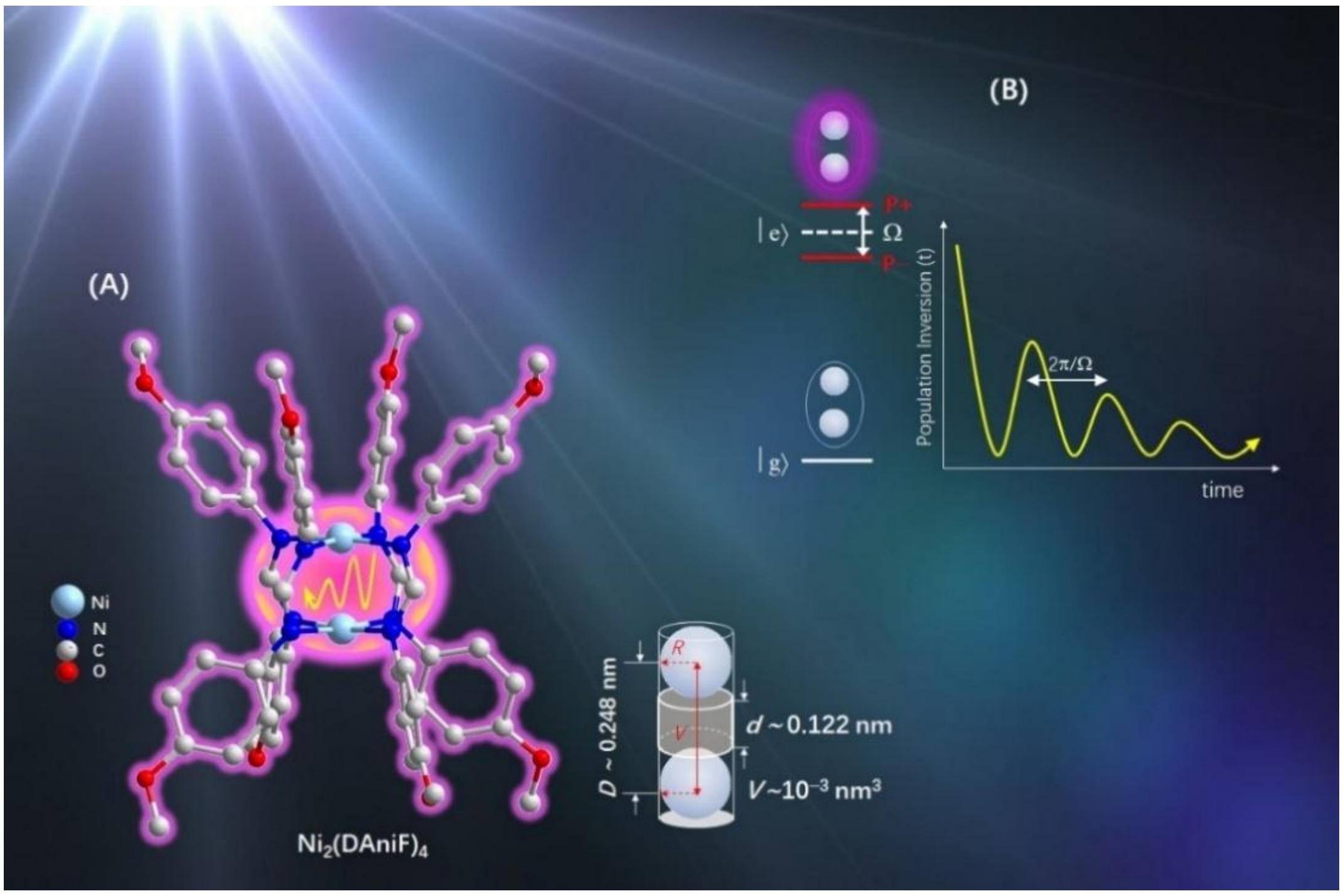

**Figure 2. Geometric structure and optical responses of the $Ni_2$ complex.** (A) The molecular structure of $Ni_2(DAniF)_4$ with the geometric dimensions of the $Ni_2$ unit shown in the inset. The inset highlights the separation between the two Ni atoms and their ionic radii, emphasizing the atomistic scale of the $Ni_2$ optical resonator. (B) Schematic representation of the Rabi oscillation for the $Ni_2$ qubit under visible light excitation. The diagram illustrates the quantum state transitions that results in vacuum Rabi splitting of the electronic normal mode in a single $Ni_2$ molecule.

**Optical Responses to Weak Incoherent Excitation.** $Ni_2(DAniF)_4$ exhibits a single, asymmetric emission at ~380 nm in the fluorescence spectra (Figure 3A) when the excitation wavelength ($\lambda_{ex}$) varies from 220 to 300 nm, showing a thermal scattering field featuring photon bunching.[6] This fluorescence state does not correspond to any electronic transition for the complex [27,28] and thus, cannot be attributed to the population decay of the excited electronic states of the molecule. This transition energy of $Ni_2$ is higher than that measured for $Cu_2$ (420 nm)[26] and $Mo_2$ (400 nm) ($Mo_2(O_2CCH_3)_4$)[30] in photoluminescence, representing different scattering characteristics for the $M_2$ complexes with varying metal nuclearities. As the excitation energy decreases from 330 nm to 400 nm, multiple emission peaks emerge (Figure 3B), indicating the single modes of the EM field continuum scattered by $Ni_2$. For each excitation, the highest energy scattering peak appears at ~380 nm, with subsequent peaks progressing towards lower energy with gradually decreased energy and intensity such that the overall spectral profile remains the same shape as that of the thermal field (Figure 3A). These results support our proposal in previous study[26]that the $Mo_2$ unit acts as an atomistic resonator operating with visible light, generating an intense local EM field that drives the light-molecule interaction. Most importantly, the scattering spectra indicate that the $M_2$ complex can trap visible light over a broad wavelength range and emit photons of discrete wavelengths. Therefore, the scattered radiation from the coupled two-level molecule, being quantized by the dinickel nonlinear medium, is expected to feature squeezed states and photon antibunching,[5,6] the two most intriguing phenomena in

quantum optics.[7]

For a single two-level transition, photon emission from the excited molecule occurs after absorption of the incident radiation in free space. Therefore, at the single-molecule level, the photon emission must occur in accordance with sub-Poissonian statistics, with the second-order correlation $g^{(2)}(0) = 0$, i.e., by photon antibunching.[5,6, 31] The phenomenon of photon antibunching is of fundamental scientific and technological interest, which has previously been demonstrated by exciting a single atom[6,32] and molecule[31,33], pumped with a laser beam. Remarkably, in this $M_2$ (including $Ni_2$ and $Mo_2$) molecular system, exciton-photon coupling occurs under irradiation with visible light in free space and under ambient conditions. The $M_2$ unit, e.g., $Ni_2$ and $Mo_2$,[30] thus behaves as a diatomic resonator, periodically absorbing incident photons through the ground state and emitting a single photon at a time from the excited state, processing the M↔M charge transfer (CT) *via* quantum tunneling.[19,20] Thus, the unprecedented optical phenomenon observed here can be justified as the excitation being coupled to the scattering field of the $Ni_2$ unit through photon antibunching, as proposed in earlier theoretical study.[34] The light scattering at ~380 nm defines the resonance energy of the two-level CT excitation ($\omega_0$) for the $Ni_2$ emitter, while quantifying the natural frequency of the $Ni_2$ diatomic resonator ($\omega_L$). Thus, the CT normal mode $\omega_0$ is intrinsically resonantly coupled to the field mode $\omega_L$.[10] This electron-photon coupling event is similar to the light-matter interaction in the superconducting circuit system,[12,24,35] where the Cooper pairs of electrons are transported across the junction gap by microwave excitation. Because the optical excitation event takes place in solution at room temperature, where the transition dipoles are randomly oriented and dynamic, the molecule-field interaction must occur at the single-molecule or few-molecule ensemble level, and independently with respect to the environment (neighboring $Ni_2$ and solvent molecules), showing the nature of the JC molecule.[5,10] Obviously, the phase matching between the individual molecular quantum systems is of particulars importance for the observation of the quantum effects in free space, which can be highly sensitive to environmental perturbations. The unique functionality of the $M_2$ resonator in trapping and converting classical light into non-classical light benefits from the enhanced spontaneous emission of the $M_2$ emitter due to the Purcell effect. Given the extremely small mode volume for the $Ni_2$ resonator, i.e., $V \approx 10^{-3}$ nm$^3$, despite a small quality factor $Q = 7$ (Figure S1), the Purcell factor is calculated to be Fp $\approx 10^{10}$ for the $Ni_2$ emitter, which is four orders of magnitude larger than that for the Au NPoM setup.[21]

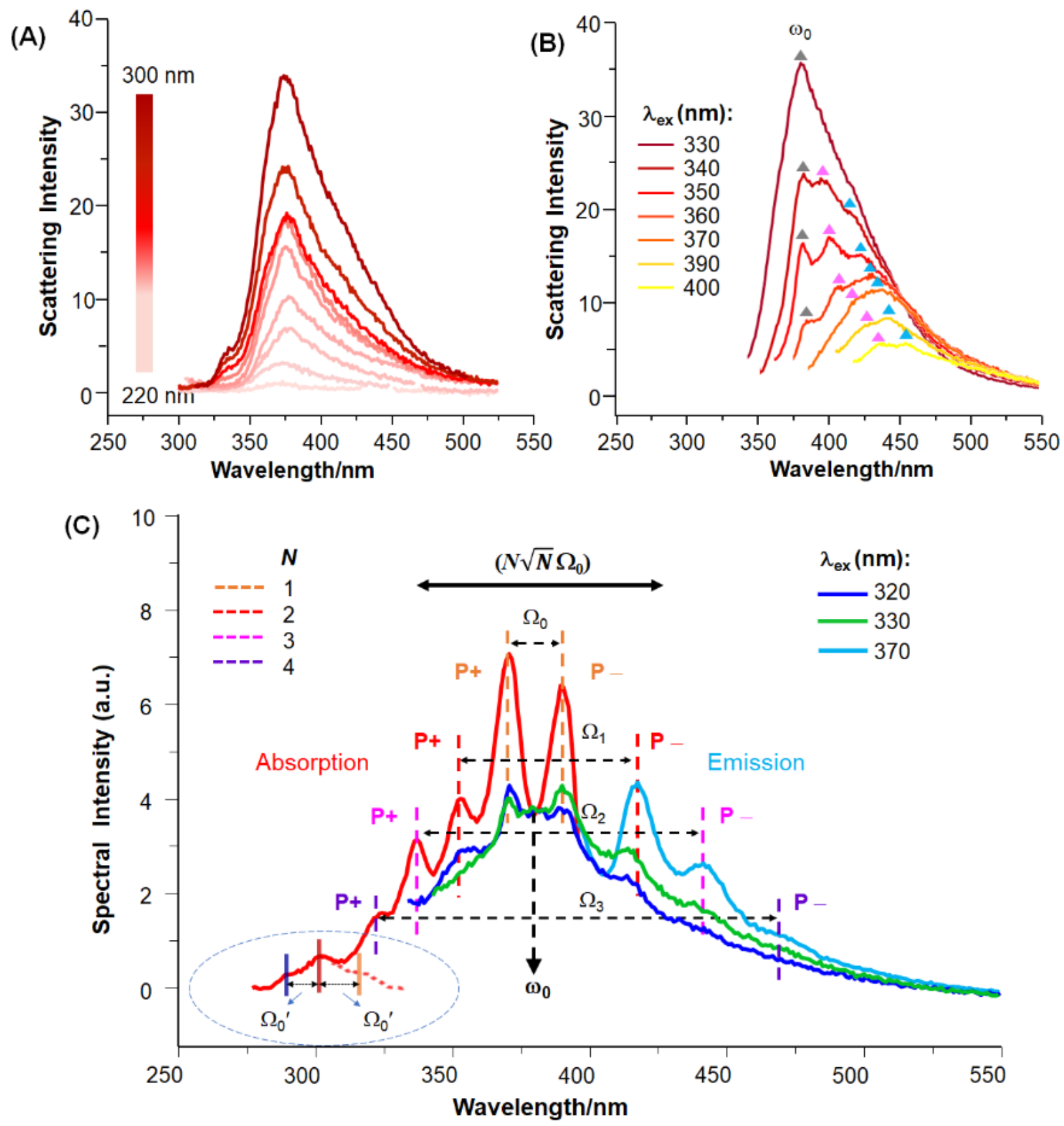

**Figure 3. Optical responses of $Ni_2(DAniF)_4$ to weak, incoherent excitations.** (A) Scattering spectra of $Ni_2(DAniF)_4$ excited at 220-300 nm, showing the thermally averaged electromagnetic field indicative of photon bunching. (B) Scattering spectra of $Ni_2(DAniF)_4$ excited at 330-400 nm, revealing discrete single modes of the optical field continuum and showing the quantization of the classical light field by the $Ni_2$ unit. (C) Excitation and fluorescence spectra displaying the Rabi splitting states for the $N$-molecule ensembles. The circled region shows the Mollow triplet at 300 nm resulting from excitation of the blue sideband ($\omega_0 + N\Omega_0$) ($N$ = 5) of the Mollow triplet at $\omega_0$.

**Resonance Fluorescence from Incoherent Excitation and Vacuum Rabi Splitting of $\omega_0$.** Excitation of a dilute dichloromethane solution of $Ni_2(DAniF)_4$ at 320 nm and 330 nm produces the photoluminescence with paired emissions symmetrically distributed around the 382 nm position, but the central peak at $\omega_0$ is quite weak or "dark" (Figure 3C). This unusual spectral feature indicates that the two-level CT excitation of $Ni_2$ ($\omega_0$) is resonantly coupled to the scattering field ($\omega_L$),[10,12] evolving discrete Rabi doublets centered at $\omega_0$ for the single molecules and the *N*-molecule ensembles as well (Figure 3C).[12] The exceptionally weak and narrow resonance peak, in a shape of Lorentzian line, can be unambiguously assigned to Rayleigh scattering of the $Ni_2$ unit, suggesting an exciton-photon coupling at the single-molecule level and by photon antibunching.[31] The symmetric spectral distribution at the resonance is interpreted as the population decay of the collectively dressed states (Figure 1B). The results are in agreement with theoretical simulations of photon-photon correlation of the scattering field for two strongly coupled two-level emitters (atom or molecule)[36], showing photon pairs of in the virtual processes and interaction-induced sidebands in the emitted spectrum. Experimentally, the light-molecule interaction observed here is somehow similar to the resonant coupling of a perovskite emitter embedded in a plasmonic cavity by incoherent excitation of a high-energy laser beam[37], suggesting that the dinickel unit plays the role of a microcavity operating under ambient conditions.

In this $Ni_2$ system, the energy levels of the dressed states for the higher order collective coupling are numerically confirmed by combing the excitation (emitted at 418 nm) and emission (excited at 370 nm) spectra, which resolve four pairs of Rabi doublets of $\omega_0$ (Figure 3C). The two sharp peaks of the inner pair, at 372 nm (26882 $cm^{-1}$) and 392 nm (25510 $cm^{-1}$), are assigned to the upper (P+) and lower (P−) exciton polaritons for the single molecules, giving a coupling strength of $\Omega_0 = 1372\ cm^{-1}$ (0.17 eV). The effective Rabi frequencies for the first, second, and third order collective coupling (Figure 1B), namely, $\Omega_1$, $\Omega_2$, and $\Omega_3$, are measured from the spectra to be 4300 $cm^{-1}$, 7000 $cm^{-1}$, and 9590 $cm^{-1}$, respectively, in excellent agreement with the calculated values of $N\sqrt{N}\Omega_0$, i.e., 3874 $cm^{-1}$ ($N = 2$), 7120 $cm^{-1}$ ($N = 3$), and 10960 $cm^{-1}$ ($N = 4$), respectively (Table S1). Remarkably nearly identical data are found for the $Mo_2$ system.[30] The relatively large deviation of $\Omega_1$ from $2\sqrt{2}\Omega_0$ may be due to the presence of the second sidebands[38] of the Mollow triplet in the vicinity of the Rabi components for the two-molecule system (*vide infra*). For the $N$ = 4 ensemble, the measured collective coupling strength reaches 1.2 eV and the normalized coupling rate = 0.18, falling in the ultrastrong coupling regime ($\eta > 0.1$).[9] The quantitative agreement between the measured and analytical collective coupling rates indicates that the $N$ molecules in an ensemble are simultaneously and individually coupled to $N$ photons of the same single field mode, demonstrating the effective JC molecular system extended from the JC molecule.[1,10,15] The scaling of the collective coupling rate for the $M_2$ systems, which differs from the $\sqrt{N}\Omega$ scaling for the Dicke and TC model, is explicitly illustrated by the magnitude $\sqrt{2}\Omega$ for the two-atom system with only one atom coupled to the field mode due to the dipole blockade.[39,40,41] The dynamics of the JC molecule gives rise to squeezed states of the EM field, [5,42]which enables breakdown of the dipole blockade in the two-atom Dicke model.[43] For decades, much effort has been devoted to the study of the additional sidebands of the Mollow triplet in resonance fluorescence for multi-atom systems.[36,38,44,45,46,47] However, multiple Rabi doublets from the extended vacuum Rabi splitting for discrete multi-emitter ensembles have not been documented.

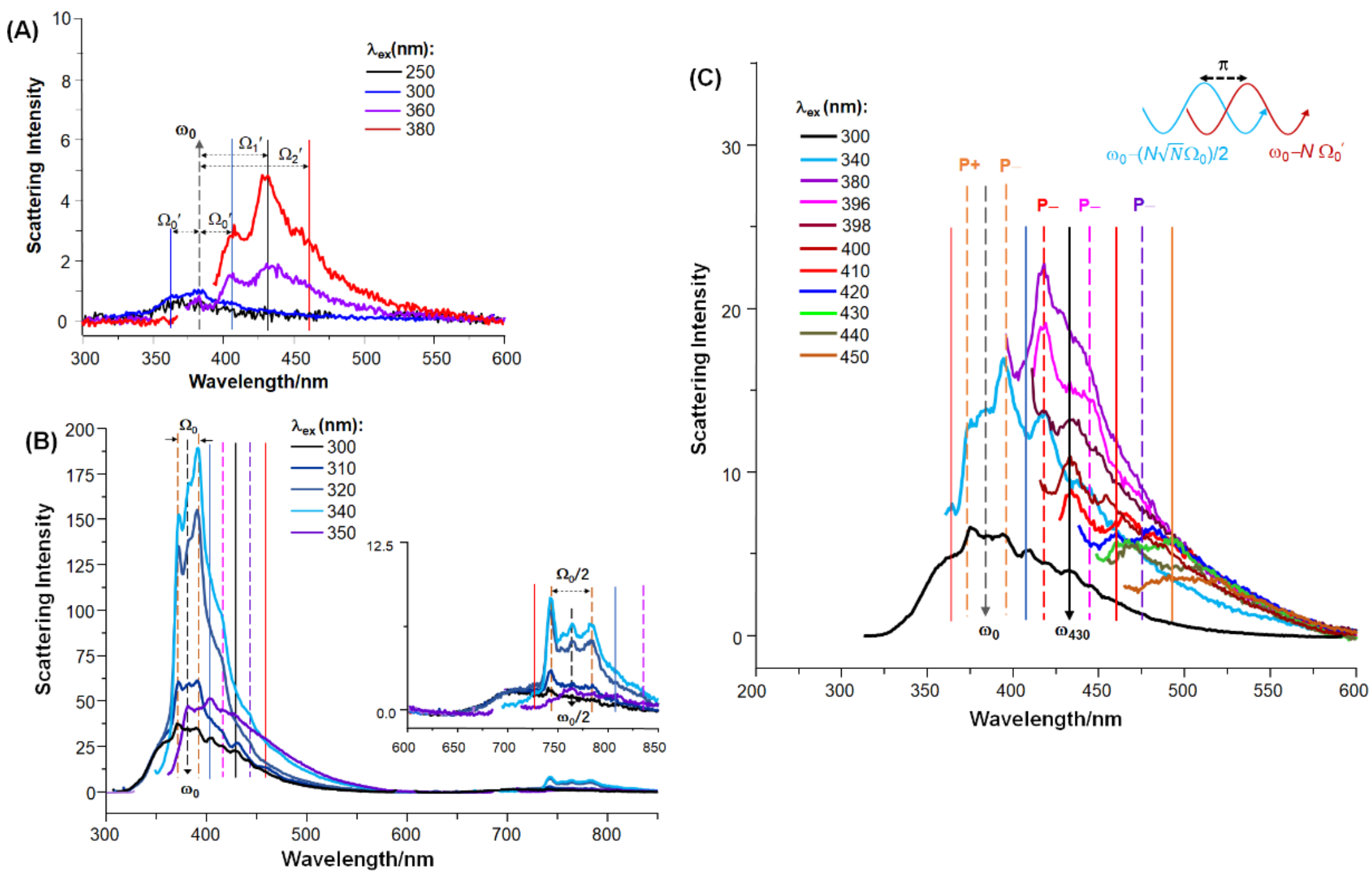

**Figure 4. Quantization of classical light through collective coupling.** (A) Photoluminescence spectra showing the low-energy sidebands of the Mollow triplet at resonance. Excitation at 380 nm produces a Mollow triplet at 430 nm, the position corresponding to the second sideband ($\omega_0 - 2\Omega_0'$). (B) Excitation with laser beams of $\lambda_{ex}$ < 380 nm produces resonance fluorescence spectra exhibiting a staircase structure of the exciton-polaritonic states. Note that the resonance peak at ~ 380 nm ($\omega_0$) is typically weak and narrow. The Rabi states and Mollow states are marked by dashed and solid vertical lines, respectively, with consistent color codes representing different orders of collective coupling, as shown in Figures 1B, 3C and 4A. The inset highlights the zoomed-in region from 600 to 800 nm, where the two peaks are separated by ~ 700 $cm^{-1}$, corresponding to half the coupling strength ($\Omega_0/2$). (C) Transformation of the Rabi doublets into the Mollow triplets occurs as the excitation energy is lowered from 380 nm to 410 nm, signaling the progression of sideband excitation in the $Ni_2$ molecular system. The inset illustrates the phase difference ( ~ 700 $cm^{-1}$) between the Rabi doublet and the Mollow triplet, highlighting the phase-sensitive transition driven by the collective coupling.

**Phase Transition of the Nonclassical Field and Quantization of Classical light.** For the concentrated solution of $Ni_2(DAniF)_4$ ($C$ = 0.1 mM), high-energy excitation (e.g., $\lambda_{ex}$ = 250 nm) produces the extremely weak $\omega_0$ peak and two sidebands that are barely visible in the fluorescence spectrum (Figure 4A, black). This triplet fluorescence is intensified by lowering the concentration to $C$ = 0.0125 mM and excitation at $\lambda_{ex}$ = 300 nm, showing three transitions occurring at 380 (26316 $cm^{-1}$), 360 nm (27778 $cm^{-1}$) and 402 nm (24876 $cm^{-1}$) (Figure 4A, blue). Therefore, this spectral profile characterizes the Mollow triplet at the resonance $\omega_0$ with two satellites at $\omega_0 \pm \Omega_0'$,[13,14] where $\Omega_0'$ = 1450 $cm^{-1}$, reasonably larger than $\omega_0$ determined from the vacuum Rabi splitting. However, this Mollow triplet structure is severely distorted by the incoherent excitation, which induces an asymmetric spectral distribution, and by the squeezed field, which broadens the peaks and reduces the central peak intensity.[48] The increase in spectral intensity with decreasing the concentration strongly supports that the molecule-field coupling is driven by photon antibunching which is diminished with decreasing the intermolecular distance.[38,49,50] Lowering the excitation energy produces three peaks at 404 nm, 430 nm, and 457 nm (Figure 4A). These peaks are assigned to the fundamental and the additional low-energy sidebands ($\omega_0 - \Omega_n'$) ($n$ = 0-2) of the three Mollow triplets at $\omega_0$, with $\Omega = \Omega_0' = 1450$ cm$^{-1}$, $\Omega_1'$ (= $2\Omega_0'$) and $\Omega_2'$ (= $3\Omega_0'$),, in agreement with the theoretical predictions for driven multi-atom systems.[13,38,45,46] Similar structures have been observed in the spectra for the $N$-atom Rb system, where pairs of the sidebands are extended outward as $N$ increases from 2 to 4.[47] Interestingly, excitation at 380 nm yields an intense Mollow triplet with a central peak at 430 nm (Figure 4A, red), indicating a shift of the resonance from $\omega_0$ to ($\omega_0 - \Omega_1'$) or ($\omega_0 - 2\Omega_0'$). The formation of the Mollow triplet at $\omega_{430}$ indicates that the second sideband ($\omega_0 - 2\Omega'$) of the Mollow triplet of the two-molecule system[38] is excited by two 430-nm photons emitted from the molecular ensemble through photon bunching.[36,50] Sideband excitation [50] is also observed in the absorption spectrum. As shown in Figure 3C, the excitation spectrum exhibits three peaks at $\omega_{300}$ and $\omega_{300} \pm \Omega_0'$, which characterizes the Mollow triplet at 300 nm, corresponding to the blue sideband of the fundamental Mollow triplet for the $N$ = 5 ensemble, i.e., ($\omega_0 + N\Omega_0'$) = 299 nm (*vide supra*).

As shown in Figures 4B and 4C, $Ni_2(DAniF)_4$ in a dilute solution (0.1 mmM) shows more intense fluorescence, which is significantly different from the spectra in the concentrated solution (0.1 mM) (Figure 4A). The fluorescence spectra generated by laser excitation at $\lambda_{ex} < 380$ nm exhibit the Rabi and Mollow states of $\omega_0$. When excited at 300 nm, the red sidebands of both the Rabi doublets and Mollow triplets appear in the spectra, with the transition energies corresponding to the data in Figures 3C and 4A. In the spectrum, the two adjacent peaks, with one from the Rabi doublets (Figure 4B, dashed vertical lines) and the other from the Mollow triplets (solid vertical lines) or *vice versa*, are separated by approximately 700 cm$^{-1}$, corresponding to $\pi$ or half of the zeroth order coupling strength $\Omega_0$. The spectra are duplicated at half the coupling energy scale, (Figure 4B, inset). It appears that the system acts as a nonlinear medium, with both input and the output involving photons emitted directly from the molecular unit itself. The emission features at half the coupling scale could signify virtual photon emission from the ground state, representing transitions enabled by the squeezed scattering field and atomistic confinement inherent to the $Ni_2$ system. Recent work has demonstrated that vacuum Rabi splitting arises from virtual excitations and their associated two-mode squeezing, observable in larger systems through collective coupling as described by the Dicke model,[51] providing a theoretical framework consistent with the spectral features observed in our system. While nonlinear optical processes such as spontaneous parametric down-conversion (SPDC) are not explicitly demonstrated here, the fluorescence features could alternatively be interpreted in this framework, supported by the strong scattering field and atomistic confinement.

When excited by a laser beam of $\lambda_{ex} \geq 380$ nm, the Rabi doublet profile is transformed into the Mollow triplet structure (Figure 4C), signaling the sideband excitations in the multi-photon processes. Transition of polariton doublet to a Stark triplet has been reported for strongly driven semiconductor microcavity with increasing field strength.[52] In this $Ni_2$ system, the spectral transformation is completed at 400 nm excitation, where the Mollow triplet states are represented as fluorescence peaks, while the P− polaritonic states correspond to the valleys (Figure 4C). The transition energies decrease continuously and constantly, forming a staircase with each step shifting by 700 $cm^{-1}$ in energy (Figure 4C). This transition from the Rabi states to the Mollow states indicates a phase transition of the scattering field (output), induced by the incoherent laser excitations (input), showing the phase difference between the Rabi and Mollow states. For example, a phase shift is indicated by the separation of 724 $cm^{-1}$ between the P− peak at 417 nm and the central peak of the $\omega_{430}$ Mollow triplet, which is exactly scaled as $\pi$ ($\Omega_0/2$) (Figure 4C, inset). The observed phase transition in the quantized scattering field of $Ni_2$ manifests the phase sensitive population decay in a squeezed field that occurs for the JC molecular system.[5,42,43]

In the transformed spectra, the Mollow triplets show the central peak at ~ 430 nm ($\omega_{430}$) and the red sideband at $\omega_{430} - \Omega_0'$, as observed in Figure 4A. Excitations with $\lambda_{ex}$ > 410 nm produce spectra featuring two progressive peaks of similar intensities. These two peaks are assigned to the red sidebands of the $\omega_{430}$ Mollow triplets, i.e., ($\omega_{430} - \Omega_0'$ and $\omega_{430}$ − $2\Omega_0'$). These results illustrate that quantization of the scattering field is realized by collectively coupling $N$ molecules in an ensemble, and the quantum light of the discrete single modes is defined by the Rabi splitting states ($\omega_0 \pm (\Omega_0 N\sqrt{N})/2$) and the Mollow states ($\omega_0 \pm \Omega_0' N$). In cavity QED, field quantization has been realized by controlled injection of photon field into a microcavity for interaction with the Rydberg atoms,[1,2] where the signal exhibits the Rabi splitting proportional to the square root of photon number ($n$), i.e., $\Omega_0\sqrt{n}$, with $n$ up to 4. Remarkably, in this $Ni_2$ system, both the Rabi and Mollow states contribute to graining of the field and these groundbreaking results are accessible with the steady-state spectroscopic methods under ambient conditions.

**Nonlinearity of the Collective Coupling and a Proposal for the Formation of Molecular Ensembles.** At varying concentrations, i.e. $C$ = 0.1, 0.05, and 0.025 mM, excited at 360 nm, $Ni_2(DAniF)_4$ shows the half-truncated Mollow triplets with a weak $\omega_0$ peak and two well-defined red sidebands at $\omega_0 - \Omega_0'$ and $\omega_0 - 2\Omega_0'$, as shown in Figure 5A. The averaged intermolecular distances ($R$) (Figure S2) are estimated to be 237 Å, 303 Å and 387 Å, respectively. The shortest distance (237 Å) for the most concentrated solution (0.1 mM) is significantly larger than the diameter of the molecule (Figure 5A, inset), i.e., 18 Å, assuming a spherical geometry for the molecule. Therefore, the intermolecular distances in these solutions are much larger than the van der Waals distances and about an order of magnitude shorter than the wavelengths of visible light, indicating strong intermolecular interactions.[38,47] The 0.1 mM solution shows that the second sideband is stronger than the first one; however, for the dilute solutions, where the $R$ increases and the dipole-dipole interaction is weakened, the intensity and transition energy at ($\omega_0 - \Omega_0'$) increase. It is also observed that the sidebands ($\omega_0 - N\Omega_0'$) ($N$ = 1 and 2) are red-shifted with increasing the concentration (Figure 5A), indicating an enhanced coupling strength. These results are in full agreement with the theoretical predictions,[38,47] demonstrating that the light-molecule interaction is driven by photon antibunching and controlled by the strength of the intermolecular interactions.[20,38,41,47] The values of $\Omega_0'$ and $\Omega_1'(= 2\Omega_0')$ increase by about 300 $cm^{-1}$ as the concentration increases from 0.0125 to 0.1 mM, showing a relatively weak concentration dependence of the collective coupling, as expected for the JC molecular system. Quantitative analysis of the spectral data shows good linear correlations of $\Omega$ ($\Omega_0'$ and $\Omega_1'$ (Figure S3)) with $\sqrt{C}$ ($R^2$= 0.96) and $C\sqrt{C}$ ($R^2$= 0.99), as shown in Figure 5B (Table S2), demonstrating the characteristic nonlinear optical behavior of the molecular ensembles[53] and the single molecules.[25] The better linearity for the plot of $\Omega$ vs. $C$ supports our model of collective coupling scaling, i.e., $\Omega \propto N\sqrt{N}$.

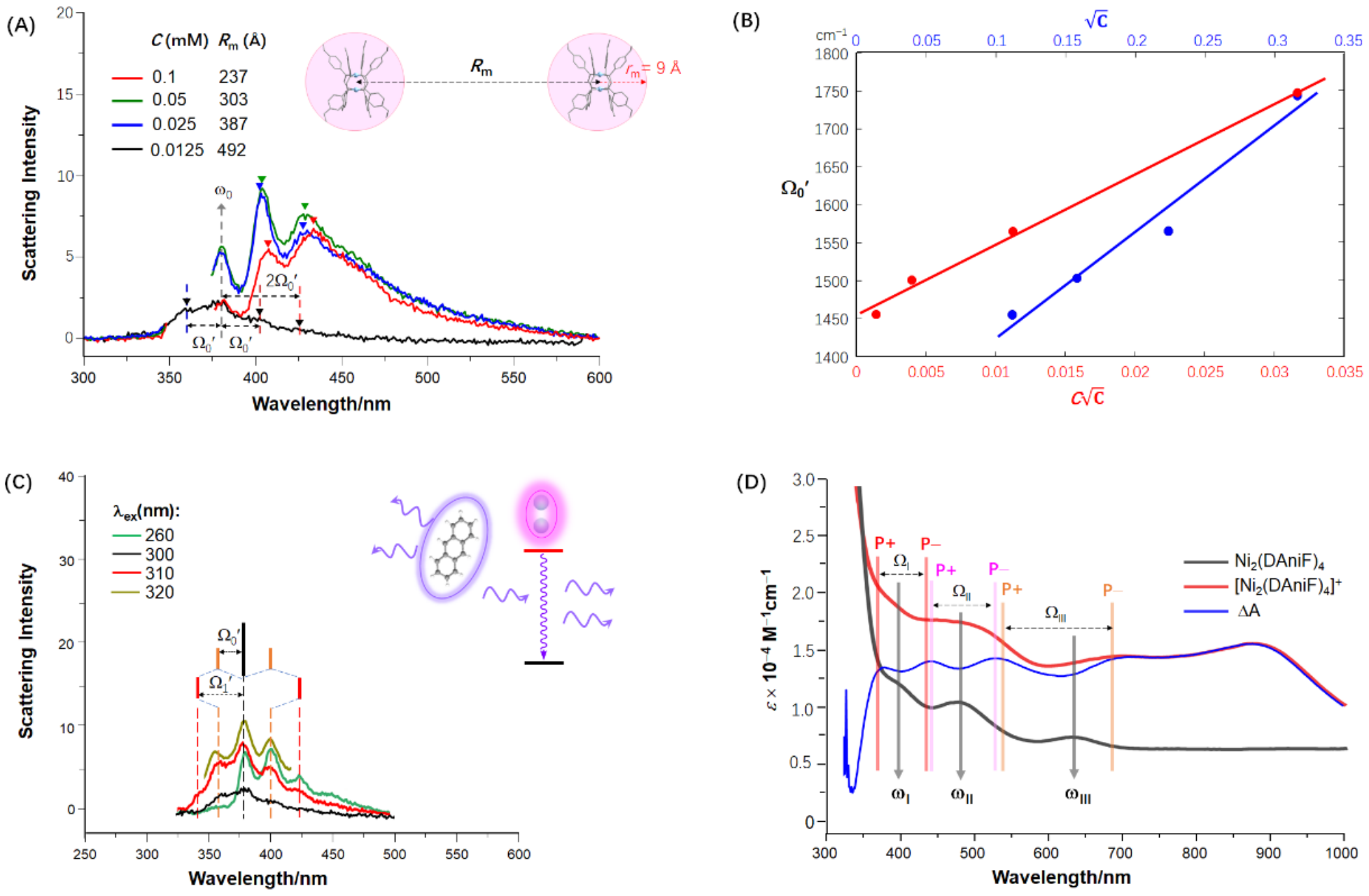

Figure 5. (A) Concentration ([*C*]) and intermolecular distance (*R*) dependence of the scattering intensity and transition energy for the $Ni_2$ system. The Mollow triplet observed for the 0.0125 mM solution is excited at 300 nm. The solutions of other concentrations, excited at 360 nm, produce the half-truncated spectra with a resonance peak ($\omega_0$) at 380 nm, showing red-shifted sidebands with increasing [*C*]. (B) Linear correlation plots of coupling strength ($\Omega_0'$) versus concentration ($C\sqrt{C}$ and $\sqrt{C}$), demonstrating the nonlinear behavior of collective coupling. (C) Mollow triplet and quintet spectra observed by introducing external, anthracene-scattered photons via stimulated emission (inset), showing the higher-order collective coupling. (D) UV-Visible spectra of the neutral (black) and oxidized (red) $Ni_2$ complexes. The difference spectra (A, blue) are derived by subtracting the absorption of the neutral complex from that of the oxidized complex,

showing the absorptions of polaritonic transitions (P+ and P) and the absorption bleaching of the electronic transition of the hybrid molecular system. The Rabi splitting bands (P+ and P−) of the excitations $\omega_I$, $\omega_{II}$ and $\omega_{III}$ are indicated by pairs of red, pink and orange vertical lines, respectively.

In addition, the fluorescence spectra (Figure 5A) show an unpredictable concentration (*C*) or distance (*R*) dependence of the scattering intensity, which requires rationalization. The range of intermolecular distances falls in the spatial region where the atoms (molecules) are dressed by zero-point field fluctuations[54], yielding the Casimir-Polder force.[55] The Casimir-Polder force is maximized when the intermolecular distance corresponds to the atomic frequencies, proportional to $R^{-7}$ for large distances.[55] A recent study has demonstrated that the Casimir effect can induce self-assembly of nanoparticle dimers and trimers in aqueous solution, which exhibit characteristic cavity modes for coherent coupling of the excitons.[56] In this $Ni_2$ system, the distinct dependence of the spectral intensity on *R* (Figure 3A) can be understood from the increase of the Casimir-Polder force with increasing the intermolecular distance towards the direction approaching the molecular transition frequency.[55] This leads us to propose that the molecular ensembles are formed by self-assembly of the molecules in solution, driven by the Casimir-Polder force.[56] An in-depth investigation of this topic is ongoing in our group.

**Enhancement of the Light-Molecule interaction by External Photons.** The proceeding discussion reveals that the light-matter interaction in this $Ni_2$ molecular system is initiated by incoherent excitation and driven by photon antibunching, and damped in the squeezed vacuum.[5,31,38,46,48] However, such a photonic environment is highly unfavorable for observation of the standard Mollow triplet evolving from the multi-photon processes,[14,38] as discussed above. Anthracene is known to emit photons from the excited state (($S_1$) to the ground state ($S_0$) at 382 nm, with the 0-0 transition energy nearly equal to $\omega_0$ of $Ni_2$. To increase the fluorescence intensity and observe the multiplets of the higher order collective coupling, we performed the fluorescence measurements of the $Ni_2$ complex mixed with a small amount of anthracene. Compared to the incoherent resonance fluorescence (Figure 5C, black) and the spectrum for pure anthracene (Figure 5C, green), the photoluminescence shows an largely enhanced Mollow triplet structure with $\lambda_{ex}$ = 320 nm and an previously unreported Mollow quintet with $\lambda_{ex}$ = 310 nm (Figure 5C). This Mollow quintet is attributed to the combination of two Mollow triplets, [44] characterized by three peaks $\omega_0$, $\omega_0 \pm \Omega_0'$ and $\omega_0$, $\omega_0 \pm 2\Omega_0'$, resulting from simultaneous excitation of the single molecules and the two-molecule ensemble,[38,45,47] respectively. These results demonstrate that the light-molecule interaction can be manipulated by introducing external photons, which is important for the real-world applications.

**Exciton-Polaritonic Transitions Observed in the Stead-State Absorption Spectra for the Cationic $Ni_2$ complex.** The neutral complex $Ni_2(DAniF)_4$ shows three absorption bands at 398 nm, 482 nm, and 635 nm in the UV-Vis spectra (Figure 5D, black), as reported in the literature.[27,28] These bands should be attributed to the two-level metal to ligand (ML) or ligand to metal (LM) CT transitions, namely, I, II and III , respectively, due to the nonbonding structure of the $Ni_2$ unit, which eliminates the vertical metal to metal electronic transition.[27,28] Notably, a significantly different spectrum is obtained for the singly oxidized complex $[Ni_2(DAniF)_4]^+$ (Figure 5D, red, Figure S4), which cannot be interpreted from the electronic transitions, but is justified on the grounds of coherent coupling of the excitation normal modes with light, as observed in $Mo_2$ systems.[26,30] In the spectrum of $[Ni_2(DAniF)_4]^+$, the resonance absorptions I, II and III, being the dark states of the bare molecule, are bleached out, as indicated by the spectral valleys in the $\Delta A$ spectrum (Figure 5D, blue).[26] The absorptions surrounding a dark state result from vacuum Rabi splitting of the two-level transition. For excitations I and II, energetically compatible photonic modes are found in the spectrum of the quantized scattering field of $Ni_2$. While the P– band at 392 nm of the single molecules (Figure 3C) is available for resonant coupling of $\omega_I$, the red Rabi and Mollow sidebands of the $N$ = 4 ensemble at 480 nm (Figure 4C) correspond in energy to $\omega_{II}$. Consequently, the transition $\omega_I$ is split into the P+ (376 nm) and P–

(440 nm) bands centered at 405 nm, giving a coupling rate $\Omega_I$ of 3870 $cm^{-1}$. For the 482 nm excitation ($\omega_{II}$), the polaritonic transitions at 442 nm (P+) and 530 nm (P−) are probed in the ΔA spectrum with a coupling rate $\Omega_{II}$ of 3740 $cm^{-1}$. The coupling strengths for these two excitations fall in range of those for the dressed organic molecules in solution in cavity-QED.[53] However, for the transition at 632 nm ($\omega_{III}$), the two surrounding absorptions, as marked by the orange vertical lines in Figure 5D, are highly asymmetric, indicating a far-off-resonant coupling. This is because the excitation is much lower in energy than the modes of the scattering field (Figure 4C), leading to coherent coupling with a large detuning, i.e., 3000 $cm^{-1}$. The observation of the exciton-polaritons of the electronic transitions for the cationic $Ni_2$ complex validates the theoretical proposal that scattered light at the resonance of a two-level atom could achieve the Autler-Townes (Rabi) splitting and the resonant optical Stark effect.[34] The absorption spectra of the $Ni_2$ complex system show that the two-level transitions of the cationic molecule are coherently coupled with a selective single mode from the quantized scattering field by the $Ni_2$ unit. Observation of the vacuum Rabi splitting of the molecular excitations in the UV-Visible spectrum provides evidence for the presence of a sequence of field modes surrounding the $Ni_2$ resonator, which is different from an optical cavity specifying only a single photonic mode. Importantly, these results, together with those from the $Mo_2$ systems,[26,30] indicate that the electronic spectra for the dimetal complexes arise from a many-body system involving electronic, photonic and possibly nuclear degrees of freedom, which cannot be interpreted from a traditional chemical perspective.

**Discussion**

Achieving resonant coupling between light and matter through frequency and phase matching has traditionally required sophisticated instrumentation and stringently controlled conditions,[1,9,11,15,22,35,53] including high-Q optical cavities, cryogenic temperatures, and high intensity laser beams. Optical experiments with systems of single or few atoms (or molecules) present significant challenges,[2,15,21,31,32,33,] especially in the ultrastrong regime.[9] In this $Ni_2$ molecular system, the light-molecule interaction is induced by weak, incoherent (or coherent) excitation of classical light, evolving the zero-dimensional exciton polaritons in free space. The resonant fluorescence spectra show that the $Ni_2$ complex molecule is an inborn emitter-resonator quantum system, which enables quantization of classical light, providing a spectrum of single mode optical fields through collective coupling. The dressed $Ni_2$ molecule, as a JC molecule,[5] is endowed with intriguing nonlinear quantum optical behaviors, such as photon antibunching and squeezed states.[6,7] This hybrid molecular system, in contrast to the Rydberg atoms in an optical cavity, breaks down the dipole blockade in the two-atom Dicke model, which maximizes the collective coupling and allows easy achievement of ultrastrong coupling. In contrast to the atomic composite system, the $Ni_2$ molecular qubit favors manipulation of the quantum system with tailorable molecular structure and tunable two-level excitations. While our study employed conventional steady-state spectroscopy techniques, the robust quantum optical effects observed in the $Ni_2$ system under these conditions demonstrate the potential for further investigation using more advanced methods, highlighting the system's applicability in diverse experimental settings. It is therefore believed that the dimetal complex molecule, as an individual molecular quantum system operating in the quantum limit with visible light, provides an excellent platform for study of quantum optics and the field effect and for testing and refinement of the related theories. The unique optical properties of the dinickel complexes hold great promise for the development of optoelectronic devices and the control of chemical reactivity. The convenient fabrication of such a molecular qubit via chemical bonding should directly benefit the construction of the quantum circuits for quantum computing and information processing in the future quantum computer.

**Data availability**

All data are available in the manuscript or the supplementary materials. The spectra raw data have been deposited at Mendeley (Mendeley Data: https://doi.org/ 10.17632/gfck77w5jm.1) and are publicly available as of the date of publication. Additional information about the data will be made available from the corresponding author upon reasonable request.

**Acknowledgments**

We acknowledge the primary financial support from the National Natural Science Foundation of China (22171107, 21971088, 21371074), Natural Science Foundation of Guangdong Province (2018A030313894), Jinan University, and the Fundamental Research Funds for the Central Universities.

**Author Contribution**

C.Y.L. conceived this project and designed the experiments and worked on the manuscript. M.M. carried out the major experimental work, including chemical synthesis, data collection and analysis, and prepared the Supplementary Information. Y.N.T., Y.L.Z. and Z.C.H. involved in the spectroscopic data analysis and assisted in manuscript preparation. J.Z. and G.Y.Z. were involved in experimental investigations.

Competing interests: The authors declare no conflict of interest.

[55] Casimir, H. B. G., Polder, D. The Influence of Retardation on the London-van der Waals Forces. Phys. Rev. 73, 360 –372 (1948).

[56] Munkhbat, B., Canales, A., Küçüköz, B., Baranov, D. G. & Shegai, T. O. Tunable self-assembled Casimir microcavities and polaritons. Nature, 597, 214-219(2021).